\documentclass[nofootinbib,twocolumn,superscriptaddress,prd,floatfix]{revtex4-1}

\usepackage{epsf,epsfig,subfigure,graphicx,amsmath,amssymb,cancel}
\usepackage{subfigure}
\usepackage{soul}
\usepackage[usenames,dvipsnames]{xcolor}

\usepackage{afterpage}
\usepackage[utf8]{inputenc}

\allowdisplaybreaks

\definecolor{DarkGreen}{rgb}{0,0.6,0}
\usepackage[colorlinks=true,linktocpage=true,linkcolor=DarkGreen,citecolor=red,urlcolor=blue]{hyperref}

\newcommand{\rmg}{\textsl{\textrm{g}}}
\newcommand{\nn}{\nonumber}
\newcommand{\bea}{\begin{eqnarray}}
\newcommand{\eea}{\end{eqnarray}}

\begin{document}
\title{1+1D Hadrons Minimize their Biparton Renyi Free Energy}
\author{Pouya Asadi}
\email{pasadi@uoregon.edu}
\affiliation{Institute for Fundamental Science and Department of Physics, \\ University of Oregon, Eugene, OR 97403, USA}
\author{Varun Vaidya}
\email{varun.vaidya@usd.edu}
\affiliation{Department of Physics, University of South Dakota, \\ Vermillion, SD 57069, USA.}

\begin{abstract}

We use a variational method to calculate the spectrum and the parton distribution function of ground state hadrons of various gauge theories in 1+1 dimensions. The template functions in our method minimize a free energy functional defined as a combination of free valence partons' kinetic energy on the lightcone and the Renyi entanglement entropy of biparton subsystems. Our results show that hadrons in these theories minimize the proposed free energy. The success of this technique motivates applying it to confining gauge theories in higher dimensions.

\end{abstract}

\maketitle

\section{Introduction}

One of the enduring questions in modern physics is a description of the spectrum of strongly coupled theories such as Quantum Chromodynamics (QCD), which describes the strong nuclear force that holds quarks and gluons together in colorless hadrons. 
Our information about a hadron structure comes from experimentally accessible structure functions such as the Parton Distribution Function (PDF), which can be isolated using factorization theorems. 
While no analytical technique exists for calculating these functions, numerical computations in certain regimes are possible with lattice QCD \cite{Ji:2013dva,Xiong:2013bka,Ji:2014gla,Ma:2014jla,Chen:2017mzz,Ma:2017pxb,Zhang:2018nsy}. However, these tools are computationally and economically expensive and do not offer any simple insight into the mechanism of strong interactions. 
 
This has motivated novel approaches (e.g. \cite{Skyrme:1961vq,Skyrme:1962vh,Schwinger:1962tp,tHooft:1973alw,Chodos:1974je,Chodos:1974pn,tHooft:1974pnl,DeGrand:1975cf,Callan:1977gz,Shifman:1978bx,Witten:1979kh,Witten:1983tx,Adkins:1983ya,Komargodski:2018odf}) for understanding confining theories, many of which were first proposed in 1+1 dimensions (1+1D). Here, gluons are not propagating degrees of freedom, there is no spin, and since the gauge coupling is dimensionful, the running effects are power suppressed. The PDF of the quarks inside hadrons of 1+1D theories is simply equal to the absolute square of their wavefunction and does not run with the energy scale, see Ref.~\cite{Asadi:2022vbl} for further details.

In this letter, we suggest that since a strongly coupled bound state is a complex system with numerous interactions between its partons, notions from complex systems, statistical mechanics, and information theory (see Ref.~\cite{PhysRev.106.620} for connections between the latter two subjects)
can be elevated to a more central role in describing its properties. 
(Notions from quantum information sciences have already been used in studying other aspects of confining gauge theories, e.g. see Refs.~\cite{Simak:1988qp,Witten:1998zw,Aharony:2003sx,Klebanov:2007ws,Bah:2007kcs,Fujita:2008zv,Kutak:2011rb,Kol:2014nqa,Wang:2014lua,Kharzeev:2017qzs,Shuryak:2017phz,Baker:2017wtt,Hagiwara:2017uaz,Liu:2018gae,Han:2018wsw,Tu:2019ouv,Arefeva:2020uec,Castorina:2020cro,Iskander:2020rkb,Jokela:2020wgs,Han:2020vjp,Kharzeev:2021yyf,daRocha:2021ntm,Baty:2021ugw,Dvali:2021ooc,Kharzeev:2021nzh,Zhang:2021hra,Hentschinski:2021aux,Dumitru:2022tud,Wang:2022noa,Benito-Calvino:2022kqa}.)
Similar to generic systems studied in classical statistical mechanics, e.g. a container of gas, that are governed by a minimum free energy principle, we conjecture partons' distribution inside hadrons are governed by a minimum free energy principle that includes a measure of entanglement between pairs of partons. 
Such a principle should not rely on the special properties of a theory, such as its symmetries, so that it can be readily applied to higher dimensional systems.

In Ref.~\cite{Asadi:2022vbl}, we tested whether the hadron wavefunction in these theories could be described as a thermal ensemble. The ansatz for the wavefunction was derived from minimizing a proposed free energy 
\bea
\label{eq:defF}
F =  E  -  T S,
\eea
where $E$ ($S$) is free parton kinetic energy on the lightcone (the von Neumann entanglement entropy of biparton subsystems).  
We observed that this description was exact in the limit of infinite parton %(bare) 
mass. The description deviates from exact numerical results as we move to lower quark masses and is no longer even approximately applicable in the deep non-perturbative regime, i.e. when $m_q/\rmg \ll 1$ with $m_q$ denoting the quark mass and $\rmg$ denoting the coupling of the confining gauge group in 1+1D. 
This motivates us to explore modifications of the free energy principle, keeping in mind that it should asymptote to the free energy of Ref.~\cite{Asadi:2022vbl} in the large quark mass limit.

Our proposal in this letter is to replace the von Neumann entropy of Eq.~\eqref{eq:defF} with either Tsallis \cite{tsallis1988possible} or Renyi \cite{renyi1961measures} entanglement entropy (both these entropy functions work for our problem); for simplicity, we use the Renyi entropy throughout this letter. This family of entropy functions finds a variety of applications in physics including quantum gravity (e.g. Refs.~\cite{Headrick:2010zt,Faulkner:2013yia,Dong:2016fnf}), cosmology \cite{Moradpour:2018ivi}, topological phases \cite{Flammia:2009axf}, non-equilibrium many body systems \cite{Rakovszky:2019oht}, and condensed matter physics \cite{Hastings:2010zka}.

We can interpret our proposal as a variational method for approximating the wavefunction and mass spectrum of the shallowest bound states of 1+1D gauge theories.

We find that our proposed free energy correctly predicts the hadrons' mass spectrum and PDF for all quark masses. The success of this approximation in reproducing the existing results motivates us to propose our minimum free energy principle as the first principle governing the properties of bound states of confining gauge groups, including those in higher dimensions.

\section{A Variational Method for Hadron Wavefunction}

We propose replacing the von Neumann entropy in the biparton free energy of Ref.~\cite{Asadi:2022vbl} by one of its generalizations, namely the Renyi entropy \cite{renyi1961measures}\footnote{One can show that for our study, Renyi and Tsallis \cite{tsallis1988possible} entropy give rise to exactly the same results. Throughout the text we work with the Renyi entropy, but all our  conclusions remain intact if we use Tsallis entropy instead.}
\bea
S_{\alpha}(\rho) = \frac{1}{1-\alpha} \ln \left( \text{Tr}\Big[\rho^{\alpha}\Big] \right),
\eea
where $\rho$ is a reduced density matrix and $\alpha$ is the \textit{order}. In the limit of $\alpha \rightarrow 1$, the Renyi entropy $S_\alpha$ reduces to the von Neumann entropy. 
We therefore propose a new \textit{biparton Renyi free energy} functional 
\bea
F_{\alpha} =  E  -  T S_{\alpha},
\label{eq:defFnew}
\eea
where $E$ denotes the kinetic energy of free valence partons on the lightcone and $S_\alpha$ is the order $\alpha$ Renyi entropy of the biparton subsystems of the hadron, see Ref.~\cite{Asadi:2022vbl} for further details. 
We propose to test whether minimizing this functional will lead to an improved description of a hadron structure.

For a meson, in 1+1D the wavefunction is described by a $q \bar q$ state since the higher Fock state contribution is suppressed for the ground state. Following Ref.~\cite{Asadi:2022vbl}, our modified free energy for the $q \bar q$ biparton system is given by 
%\small
\begin{eqnarray}
F_{\alpha}=\frac{m_q^2}{P^-} \int  & dx &  |\phi(x)|^2\left( \frac{1}{x} +\frac{1}{1-x} \right) \nonumber \\
&& \nonumber \\
\label{eq:Fmeson}
& - &  \frac{T}{1-\alpha} \ln \left( \int dx   |\phi(x)|^{2\alpha} \right),
\end{eqnarray}
where $P^-$ is the large lightcone momentum, $\phi(x)$ is the meson wavefunction, $x$ is the momentum fraction of $P^-$ carried by the quark, and $T$ is a Lagrange multiplier. 
This functional is minimized by the following ansatz
\begin{eqnarray}
\label{eq:Ansatz}
 |\phi(x)|^2 = \left( \frac{m_q^2}{\mathcal{T}^2} \right)^{\frac{1}{\alpha-1}} \Big[x(1-x)\Big]^{\frac{1}{1-\alpha}},
\end{eqnarray}
where 
\begin{eqnarray}
\mathcal{T}^2 = T P^- \frac{ \alpha}{(1-\alpha)\int dx   |\phi(x)|^{2\alpha} },
\label{eq:defT}
\end{eqnarray}
whose numerical value will be determined by the normalization condition on the PDF.

Similarly, following the prescription from Ref.~\cite{Asadi:2022vbl}, after minimizing the biparton Renyi free energy of a baryon, we arrive at the ansatz
\begin{widetext}
 \begin{eqnarray}
 &&|\phi(x, z_1, z_2, \cdots, z_{N-2})|^{2} =\left( \frac{m_q^2}{\mathcal{T}^2}\right)^{\frac{1}{\alpha-1}} \left( x \Bigg[ 1-x-\sum_{k=1}^{N-2} z_k\Bigg ] \prod_{k=1}^{N-2} z_k  \right)^{\frac{1}{1-\alpha}} ,
 \end{eqnarray}
where $N$ is number of colors and again $\mathcal{T}$ will be determined numerically by demanding the right normalization for the wavefunction. We can now write an expression for the quark PDF inside a baryon (see Ref.~\cite{Asadi:2022vbl} for further details) 
\bea
    f_q (x) =N \left( \frac{m_q^2}{\mathcal{T}^2}\right)^{\frac{1}{\alpha-1}} {\displaystyle \left( \prod_{i=1}^{N-2}  \int_0^{1-x-\sum_{j=1}^{i-1} z_j}  dz_i \right) } ~ \left( x \Bigg[ 1-x-\sum_{k=1}^{N-2} z_k\Bigg ] \prod_{k=1}^{N-2} z_k  \right)^{\frac{1}{1-\alpha}}  .
    \label{eq:ansatzNlarger}
\eea
\end{widetext}

We will use Eqs.~\eqref{eq:Ansatz} and \eqref{eq:ansatzNlarger} as the template functions in a variational method for finding eigenvalues and eigenfunctions of ground state hadrons in various confining theories in 1+1D. The variational parameter is the order $\alpha$ and will be calculated numerically by demanding that the wavefunction minimizes the expectation value of the Hamiltonian. The specifics of a particular theory will therefore be reflected in how $\alpha$ varies as a function of $m_q$. We can now look at specific theories in 1+1D and carry out this calculation. Since the gauge coupling $\rmg$ is dimensionful in 1+1D theories, we normalize all other dimensionful quantities by $\rmg$. 

\section{Schwinger model}

This is a U(1) gauge theory in 1+1D with a  fermion of mass $m_q$ coupled to a photon \cite{Schwinger:1962tp}. 
The photon can be eliminated using gauge redundancy and equations of motion, i.e. it is not a propagating mode in 1+1D.  
We work in the Infinite Momentum Frame (IMF) with $P^- \rightarrow \infty$, where $P^+$ acts as the Hamiltonian. 
In this frame the right-handed component of the fermion field, $\psi_R$, is not a propagating degree of freedom either and can be eliminated using equations of motion. 
Having integrated out these fields, we find an effective four-fermion interaction term in the Hamiltonian written in position space as \cite{Coleman:1974bu,Coleman:1985rnk} 
\bea
H_{\text{int}}(x^0) = \rmg^2\int dx^1 dy^1 J(x^0,x^1)|x^1-y^1|J(x^0,y^1),
\eea
where $J$ is the bilinear left handed fermion current $ \psi_L^{\dagger} \psi_L$.
The expectation value of the Hamiltonian for the ground state meson in this model is given by
\small
\begin{eqnarray}
\label{eq:Hsch}
&& M_{\mathrm{hadron}}^2 =\langle \phi| P^{-} P^+ |\phi \rangle =\langle \phi| P^{-} (H^0+H_{\text{int}}) |\phi \rangle \\
&=&m_q^2\int dx |\phi(x)|^2\left(\frac{1}{x}+\frac{1}{1-x}\right)+\frac{\rmg^2}{\pi}\int dx dy \phi(x) \phi^*(y) \nn \\
&+& \frac{\rmg^2}{2\pi} \int dx dy \frac{|\phi(x)-\phi(y)|^2}{(x-y)^2}. \nn
\end{eqnarray}
\normalsize
Here $M_{\mathrm{hadron}}^2$ is the invariant mass squared of the meson bound state. We now plug in the ansatz from Eq.~\eqref{eq:Ansatz} and minimize $M_{\mathrm{hadron}}^2$ over $\alpha$ to find the mass and wavefunction of the meson. 
\begin{figure}
\centering
\includegraphics[width=0.9\linewidth]{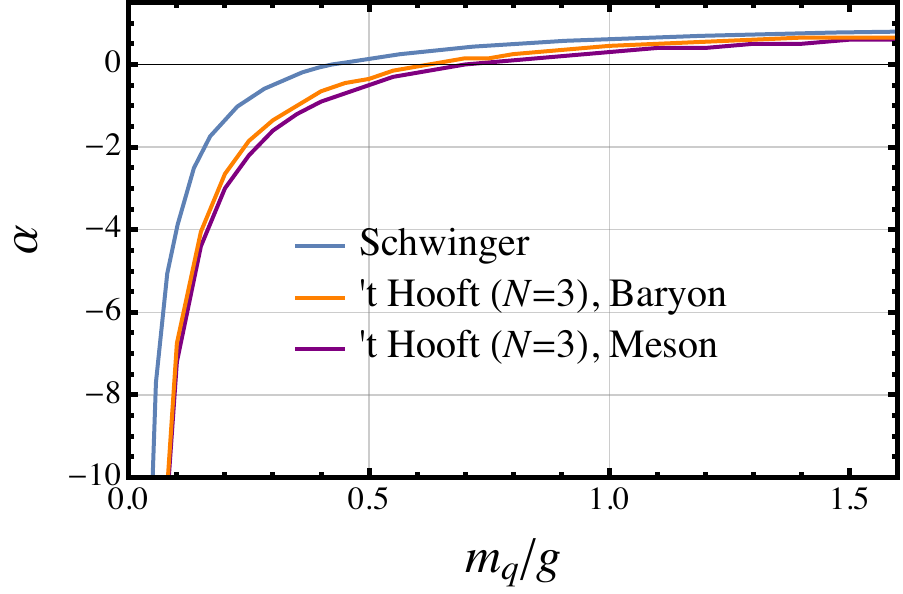}
\caption{ The order $\alpha$ for the Schwinger model (blue), mesons of the 't Hooft model with $N=3$ (purple),  and baryons (orange) of the 't Hooft model with $N=3$. We find that at large $m_q/\rmg$ values $\alpha \rightarrow 1$, i.e. our biparton Renyi free energy approaches the thermal free energy of Ref.~\cite{Asadi:2022vbl}. }
\label{fig:Alp}
\end{figure} 

We show the value of order $\alpha$ for different quark masses in Fig. \ref{fig:Alp}. The order is a monotonically increasing function of $m_q$, asymptoting to $-\infty$ as $m_q\rightarrow 0$ and approaching 1 as $m_q \rightarrow \infty$. What is peculiar is the negative order at low quark masses. 
We observe that the shift into negative order happens as we transition into the non-perturbative ($m_q \ll \rmg$) regime. 
\begin{figure}
\centering
\includegraphics[width=
0.95\linewidth]{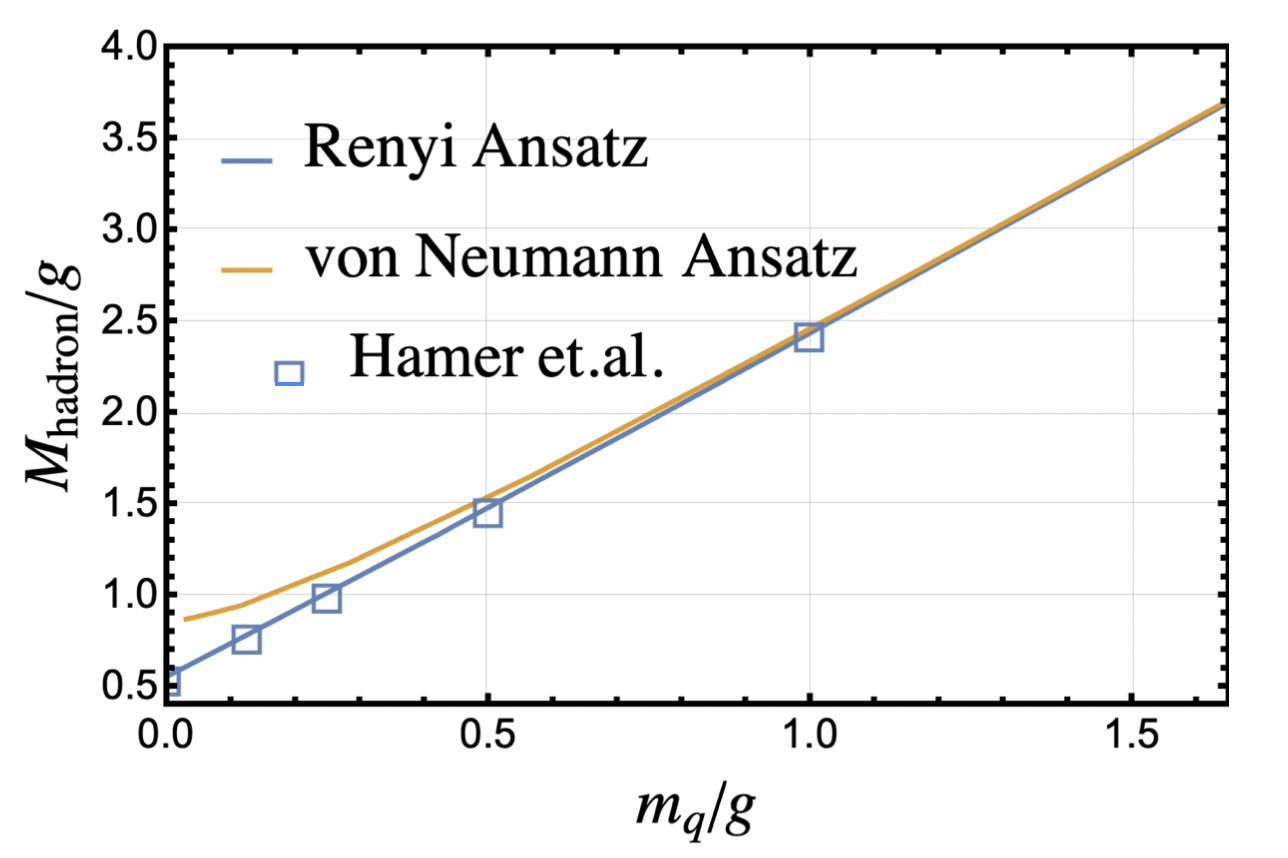}
\caption{Our prediction for the ground state meson mass in the Schwinger model (solid blue) that agrees perfectly with the existing results (blue squares) \cite{Sriganesh:1999ws}. In the $m_q/\rmg \gg 1$ limit, $\alpha \rightarrow 1$ and our result converges to the thermal description of Ref.~\cite{Asadi:2022vbl} that used the von Neumann entropy instead of the Renyi entropy (golden).
}
\label{fig:SchS}
\end{figure} 
\begin{figure}
\centering
\includegraphics[width=
0.95\linewidth]{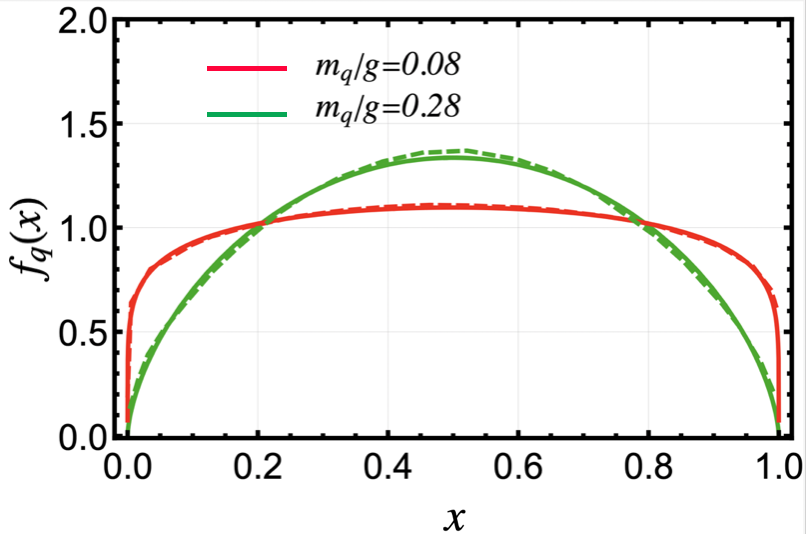}
 \caption{PDF of the Schwinger model ground state meson for two different values of quark mass. Our results (solid) agree perfectly with the existing results (dashed) \cite{Sriganesh:1999ws}.}
 \label{fig:SchPDF}
 \end{figure}

Putting the calculated value of $\alpha$ into the ansatz of Eq.~\eqref{eq:Ansatz}, we find the meson mass (Fig.~\ref{fig:SchS}) and PDF (Fig.~\ref{fig:SchPDF}) for different values of $m_q$. 
We find excellent agreement for all values of quark masses (compared to Ref.~\cite{Sriganesh:1999ws}) which supports our free energy conjecture.

%\section{'\makeLowercase{t} Hooft model}
\section{'t Hooft model}

Non-abelian gauge groups SU($N$) in 1+1D are collectively referred to as the 't Hooft model, in recognition of 't Hooft's contribution in studying their properties in the large $N$ limit \cite{tHooft:1973alw,tHooft:1974pnl}. The Hamiltonian of the bound states in the lightcone is derived and, by solving a time-independent Schr\"{o}dinger equation, the hadron mass spectrum (as a function of the quark mass) and the wavefunction of the bound state are calculated \cite{tHooft:1973alw,tHooft:1974pnl,Hornbostel:1988fb}. 

For the rest of this work, we focus on $N=3$; extending our results to higher values of $N$ is straightforward. For one flavor of quarks in the fundamental representation of SU($N$), the expectation value of the Hamiltonian of the meson state is 
\small
%\begin{widetext}
\begin{eqnarray}
\label{eq:Hmeson}
 M_{\mathrm{hadron}}^2 &=&m_q^2\int dx |\phi(x)|^2\left(\frac{1}{x}+\frac{1}{1-x}\right)   \\
&+&\frac{\rmg^2}{2\pi} \frac{3^2-1}{3}  \int dx dy \frac{|\phi(x)-\phi(y)|^2}{(x-y)^2}, \nn
\end{eqnarray}
\normalsize
while for the baryon it is 
\small
\begin{eqnarray}
& &M _{\mathrm{hadron}}^2 =m_q^2\int dx \int_0^{1-x} dz |\phi(x,z)|^2\left(\frac{1}{x}+\frac{1}{1-x-z} + \frac{1}{z} \right) \nn \\
\label{eq:Hbaryon}
&+&\frac{\rmg^2}{2\pi} \frac{3}{2} \frac{3^2-1}{3} \int dz \int_{0}^{1-z} dx dy \frac{|\phi(x,z)-\phi(y,z)|^2}{(x-y)^2}. 
\end{eqnarray}
\normalsize
Here $\phi (x,z)$ is the joint wavefunction of the three valence partons.

We now put the template ansatz from Eq.~\eqref{eq:Ansatz} or \eqref{eq:ansatzNlarger} in Eqs.~\eqref{eq:Hmeson}--\eqref{eq:Hbaryon} and minimize the eigenvalue as a function of the order $\alpha$. The resulting $\alpha$, as a function of the quark mass, is shown in Fig.~\ref{fig:Alp}. 
Similar to the case of the Schwinger model, we find that $\alpha$ grows monotonically with $m_q/\rmg$ and in the limit of $m_q \gg \rmg$ it asymptotes to 1. Note that $\alpha\rightarrow 1$ corresponds to the von Neumann entropy, thus in this limit our free energy functional becomes identical to the free energy used in Ref.~\cite{Asadi:2022vbl}. 
\begin{figure}
\centering
\resizebox{0.85\columnwidth}{!}{
\includegraphics[width=0.9\linewidth]{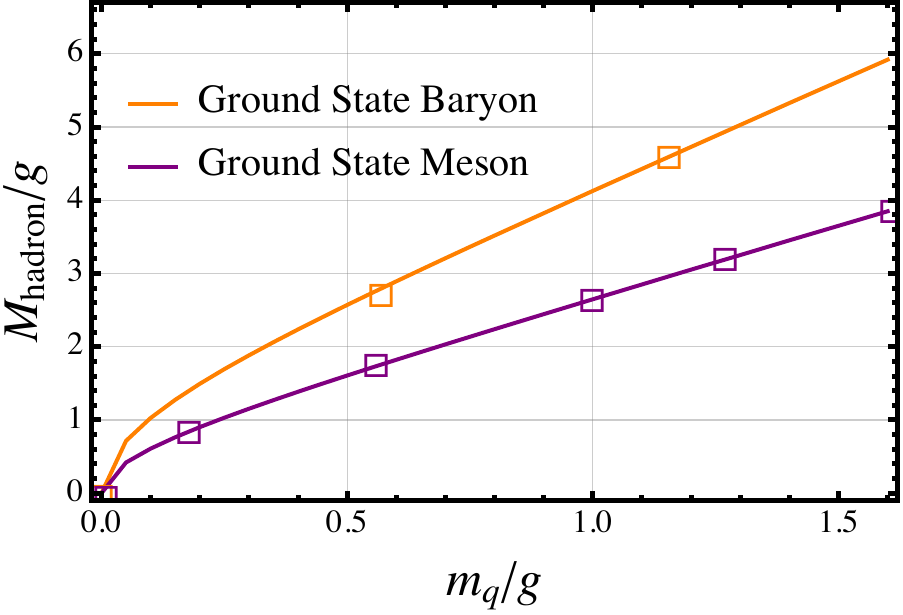}
}
\caption{Our prediction for the ground state meson (purple) and baryon (orange) mass as a function of the quark mass $m_q$ for $N=3$ 't Hooft model. We find perfect agreement between our results and existing ones from solving the lightcone Schr\"{o}dinger equations (denoted by $\square$) for both mesons \cite{tHooft:1973alw} and baryons \cite{Hornbostel:1988fb}. }
\label{fig:spectrum_thooft}
\end{figure}  

\begin{figure}
\centering
\resizebox{0.85\columnwidth}{!}{
\includegraphics[width=0.9\linewidth]{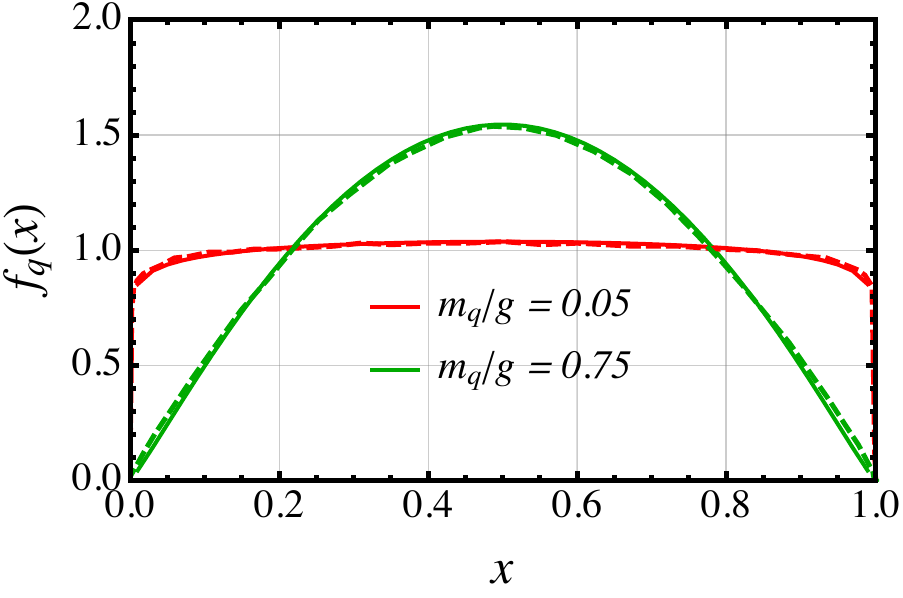}
}\\
\hspace{0.5in}
\resizebox{0.85\columnwidth}{!}{
\includegraphics[width=0.9\linewidth]{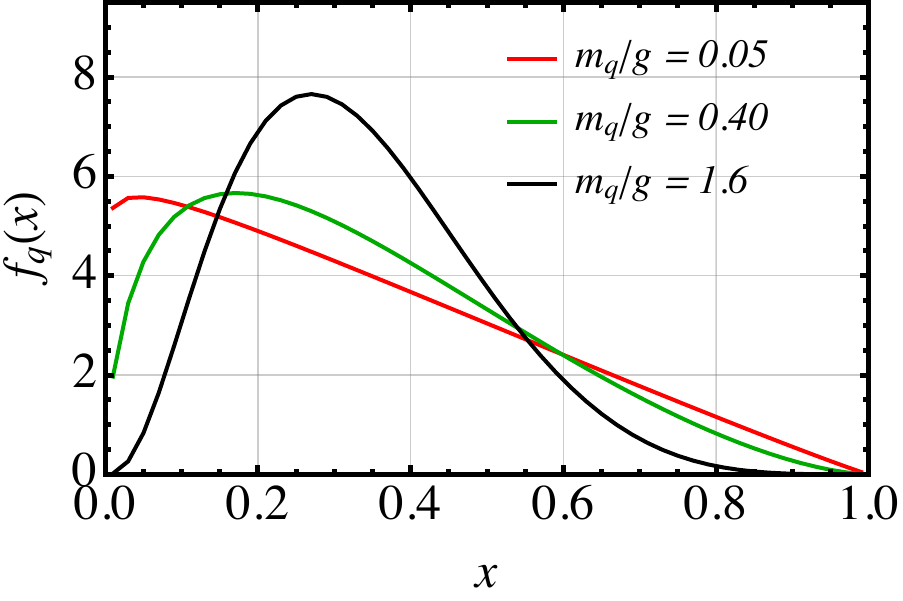}
}
\caption{\textbf{Top:} The PDF of the ground state meson of 't Hooft model with $N=3$ and for a single flavor of quark in the fundamental representation for different quark masses. Our results (solid lines) are in agreement with the existing results, see Ref.~\cite{Jia:2017uul}.
\textbf{Bottom:} PDF of the ground state baryon in the same theory for a few different quark masses. 
In the $m_q \rightarrow 0$ limit, we approach the analytic results of Ref.~\cite{Hornbostel:1988fb}. 
We are not aware of any numerical calculation of these PDFs for finite masses.}
\label{fig:thooft_pdf}
\end{figure}

We can now use these values of $\alpha$ and calculate our approximation for the mass spectrum of the ground state hadrons for different quark masses, see Fig.~\ref{fig:spectrum_thooft}. 
We are not aware of numerically accurate results for baryons in the low mass quark limit; our findings are in perfect agreement with the existing results. 
We also show the PDF of the ground state mesons (baryons) for different values of $m_q/\rmg$ on the top (bottom) plot in Fig.~\ref{fig:thooft_pdf}. 
For the case of mesons we again find perfect agreement with the results from exactly solving the lightcone Schr\"{o}dinger equation, further corroborating our biparton Renyi free energy conjecture. 
We are not aware of any existing results in the literature for baryons. 

\section{Discussion}

In this letter, we have proposed a variational method for calculating ground state wavefunction and spectrum of hadrons in 1+1D. We derive template functions from a physically motivated free energy functional made of free partons' kinetic energy in the lightcone frame and the Tsallis/Renyi entanglement entropy of  fixed momentum biparton subsystems in the bound state. 
The variation parameter is the order of the entropy and is calculated by minimizing the expectation value of the Hamiltonian of a given theory. 
We found that our method correctly reproduces the existing results for the mass spectrum and for the PDF of hadrons in the Schwinger and the 't Hooft (with $N=3$ and one flavor of quarks in the fundamental representation) model.

Our results show that hadrons in these models minimize our proposed biparton Renyi free energy functional. 
We also observe that, irrespective of the specific theory or bound state, the order of the entropy increases monotonically with the bare parton mass from $\alpha \rightarrow -\infty$ at zero quark mass to $\alpha\rightarrow 1$ at infinite quark mass limit. 
We find that the order becomes negative at low parton masses just as we enter the non-perturbative regime. Statistically, we can understand this as a switch
from maximizing high probability configurations to minimizing those with low probability. We do not yet have any deeper physical intuition about $\alpha$ beyond this interpretation.

Given our construction, we conjecture that in a biparton subsystem (in \textit{any} bound state of \textit{any} confining theory) carrying a fixed total momentum, a single parton reduced density matrix is effectively described by an ensemble which is maximally entangled at zero mass and which approaches a thermal state at large masses.

Our method can be extended to other 1+1D models. Our ultimate goal is to extend our method to models in higher dimensions, where a host of new complications (e.g. renormalization group evolution, spin, and gluon degrees of freedom) need to be accommodated. 
We leave such studies, and many other phenomenology explorations, for future work.

\section*{Acknowledgement}
We thank Chris Akers and Tim Cohen for helpful discussions. 
The work of PA is supported in part by the U.S. Department of Energy under Grant Number DE-SC0011640. 
VV is supported by startup funds from the University of South Dakota.

\bibliography{bib}
\end{document}